\documentclass[12pt]{article}
\usepackage{graphicx}
\usepackage{amsmath}
\usepackage{amssymb}
\usepackage{a4wide}

\begin{document}

\markboth{Engel and Trebin}{Structural complexity in monodisperse systems of
  isotropic particles}

\title{Structural complexity in monodisperse systems of isotropic particles}

\author{Michael Engel\footnote{Author for correspondence. Email:
    mengel@itap.uni-stuttgart.de}~ and Hans-Rainer Trebin\\
    {\normalsize Institut f\"ur Theoretische und Angewandte Physik,}\\
    {\normalsize Universit\"at Stuttgart,}\\
    {\normalsize Pfaffenwaldring 57, D-70550 Stuttgart, Germany}\\
}

\maketitle

\begin{abstract}

It has recently been shown that identical, isotropic particles can form
complex crystals and quasicrystals. In order to understand the relation
between the particle interaction and the structure, which it stabilizes, the
phase behavior of a class of two-scale potentials is studied.  In two
dimensions, the phase diagram features many phases previously observed in
experiment and simulation. The three-dimensional system includes the sigma
phase with 30 particles per unit cell, not grown in simulations before, and an
amorphous state, which we found impossible to crystallize in molecular
dynamics. We suggest that the appearance of structural complexity in
monodisperse systems is related to competing nearest neighbor distances and
discuss implications of our result for the self-assembly of macromolecules.

\end{abstract}

\section{Introduction}

When kept long enough at low temperatures, most systems develop long-range
order. The easiest crystallization is expected for monodisperse systems,
because only topological, but no additional chemical ordering is necessary. If
the particles are isotropic, then a first reasoning suggests a preference to
form simple, close-packed crystals, since this allows all of them to have the
same, high first coordination numbers. A look at the periodic table reveals
that indeed the ground states of most metals are bcc, fcc, or hcp
\cite{Donohue1982}. Similar simple crystals are found in mesoscopic or
macroscopic systems like for example globular proteins \cite{Rosenberger1996},
monodispersed colloids \cite{Xia2000}, and bubble rafts on liquid surfaces
\cite{Bragg1947}.

In some of the systems above, quantum mechanics does not play an important
role. Hence, it should in principle be possible to understand crystallization
by using classical pair interactions. For monodisperse systems there is only
one type of interaction given by the shape of the potential function. Many
common potentials are smooth and have a single minimum only. Simulations with
these potentials usually lead to simple crystals. An example is the
Lennard-Jones (LJ) potential.

Nevertheless, the situation is not always that easy. Over the last years, more
complex structures have been observed in experiments and simulations.  A large
part of work in this direction started after the discovery of quasicrystals in
metallic alloys and was carried out with the aim to understand their
formation. Another type of complex order are periodic crystals with large unit
cells, so-called complex crystals. Whereas the lattice constants of a simple
crystal are comparable to the range of the particle interactions, the unit
cell of a complex crystal is stabilized indirectly, e.g.\ by geometric
constraints.

On the theoretical side, the obvious procedure to promote structural
complexity is to use potentials with a more complicated radial dependence. It
has been shown with double-minima potentials \cite{Olami1990}, oscillating
potentials \cite{Smith1991}, and a repulsive barrier \cite{Denton1998} that
the energy of an icosahedral phase can be lower than the energy of a class of
trial structures including close-packed phases. Surprisingly, even in the LJ
system, a quasicrystal is unstable only by a small energy difference
\cite{Roth1990}.

These early works relied on general arguments and did not try to observe real
crystals in simulations.  Dzugutov was the first to actually grow a stable
one-component dodecagonal quasicrystal from the melt \cite{Dzugutov1993},
although it was later found to be only metastable \cite{Roth2000}. He used a
LJ potential with an additional bump to disfavor the formation of simple
crystals. Although no further work on other three-dimensional systems has been
reported, there are many investigations focusing on two-dimensional systems,
where computation and visualization is easier. The first such study applied a
variation of the Dzugutov potential and found a planar dodecagonal
quasicrystal \cite{Quandt1999}. Furthermore, it was pointed out that a
decagonal quasicrystal appears with a square-well \cite{Skibinsky1999} and a
ramp potential \cite{Jagla1998}.

Experiments show that nature is quite ingenious in her way to assemble
identical particles. First of all, a few complex ground states of elemental
metals are known to exist. A notable example is $\alpha$-manganese, which has
cubic symmetry with 58 atoms per unit cell. Thermodynamically stable
high-temperature phases are $\beta$-boron with 105 atoms and $\beta$-uranium
with 30 atoms per unit cell \cite{Donohue1982}. The latter is isostructural to
$\sigma$-CrFe and known as the sigma phase. Furthermore, commensurately
modulated phases are common at high pressures \cite{McMahon2006}.

Recently, different kinds of macromolecules have been observed to
self-assemble into complex phases: (i)~Under appropriate experimental
conditions, tree-like molecules (dendrons) forming spherical micelles arrange
to a dodecagonal quasicrystal \cite{Zeng2004}. (ii)~T-shaped molecules can be
designed in such a way that they organize into liquid crystalline honeycombs
\cite{Chen2005}. (iii)~ABC-star polymers form cylindrical columns according to
square-triangle quasicrystals, a two-dimensional version of the sigma phase
\cite{Hayashida2007}, and other Archimedean tilings.

How can structural complexity be understood from bottom-up? It is instructive
to study the ground state for small portions of the system -- more or less
spherical clusters -- as a function of the particle number $N$. The structure
of small clusters ($N<200$ in three dimensions, $N<30$ in two dimensions) is
often different from the bulk crystal \cite{Doye2003}. The reason is the
competition between local lowest energy configurations and the necessity of
continuation in space. For example in a monodisperse LJ system, icosahedral
coordination occurs in small clusters \cite{Northby1987}, although the hcp
phase is the lowest energy bulk crystal \cite{Stillinger2001}. On the other
hand, if small clusters already have simple structure, then the bulk phase
will also be simple. A local order, which is incompatible with periodicity, is
a necessary condition for structural complexity in the bulk.

\section{Phase diagram for the two-dimensional system}

There are two mechanisms to introduce structural complexity: either by
destabilizing simple phases or by stabilizing a complex phase. An example for
the destabilization mechanism is the Dzugutov potential. We adopt the
stabilization mechanism, because it has the advantage that the choice of the
target structure can be controlled more directly.

The particles are assumed to interact with an isotropic two-scale
potential. A simple ansatz is the Lennard-Jones-Gauss (LJG) potential
\begin{equation}\label{eq:1}
V(r)=\frac{1}{r^{12}}-\frac{2}{r^6}-
\epsilon\exp\left(-\frac{(r-r_0)^2}{2\sigma^2}\right),
\end{equation}
which has for small $r_0$ the shape of a shoulder, and otherwise represents a
double-well with first minimum at $r\approx 1$ with depth $V(1)\approx 1$ and
second minimum around $r_0$ with $V(r_0)\approx \epsilon$. The parameter
$\sigma$ specifies the width of the second minimum.  The shoulder form has
been used extensively to observe liquid-state anomalies
\cite{SadrLahijany1998} and for understanding liquid-liquid transitions
\cite{Franzese2001}. Surprisingly, almost nothing has been known until
recently \cite{Rechtsman2006, Engel2007} about the solid-state behavior of
(\ref{eq:1}).

\begin{figure}
\centering
\includegraphics[width=0.5\textwidth]{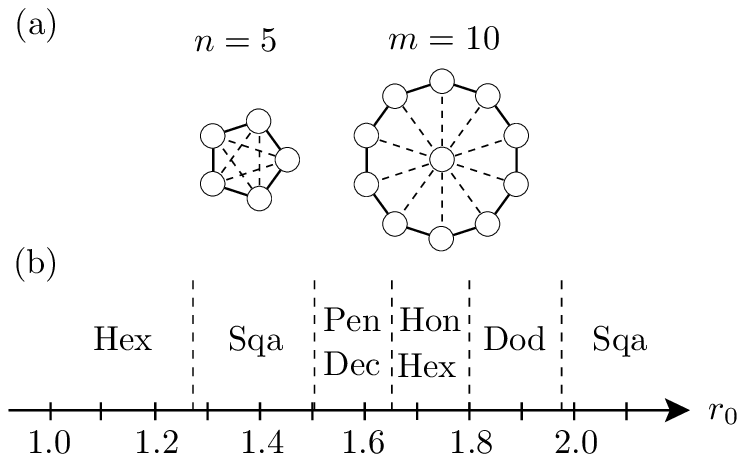}
\caption{(a)~The local lowest energy configurations in two-dimensional
  monodisperse systems are polygons with or without a central
  atom. (b)~Expected phase behavior of two-scale potentials. $r_0$ is the
  position of the second minimum.}
\label{fig:1}
\end{figure}

In two dimensions, the local lowest energy configurations are regular
polygons. As shown in Fig.~\ref{fig:1}(a), there are two possibilities:
polygons with or without central atom. An $n$-gon without central atom is
stabilized by having the second minimum at
\begin{equation}\label{eq:2}
r_0=2\cos(\pi/n).
\end{equation}
The $m$-gon with central atom is favorable for
\begin{equation}\label{eq:3}
r_0^{-1}=2\sin(\pi/m).
\end{equation}
The values $n=3$ and $m=6$ lead to hexagonal local order (Hex), the value
$n=4$ to squares (Sqa), and $n=6$ to honeycombs (Hon). Furthermore, pentagonal
($n=5$, Pen), decagonal ($m=10$, Dec), and dodecagonal ($m=12$, Dod) local
order is possible. We find that $m$-gons are not stable for other values of
$m$. For $n>6$ the vacancy is too big. The sequence of local orders as a
function of $r_0$ is expected to look like in Fig.~\ref{fig:1}(b).

\begin{figure}
\centering
\includegraphics[width=0.53\textwidth]{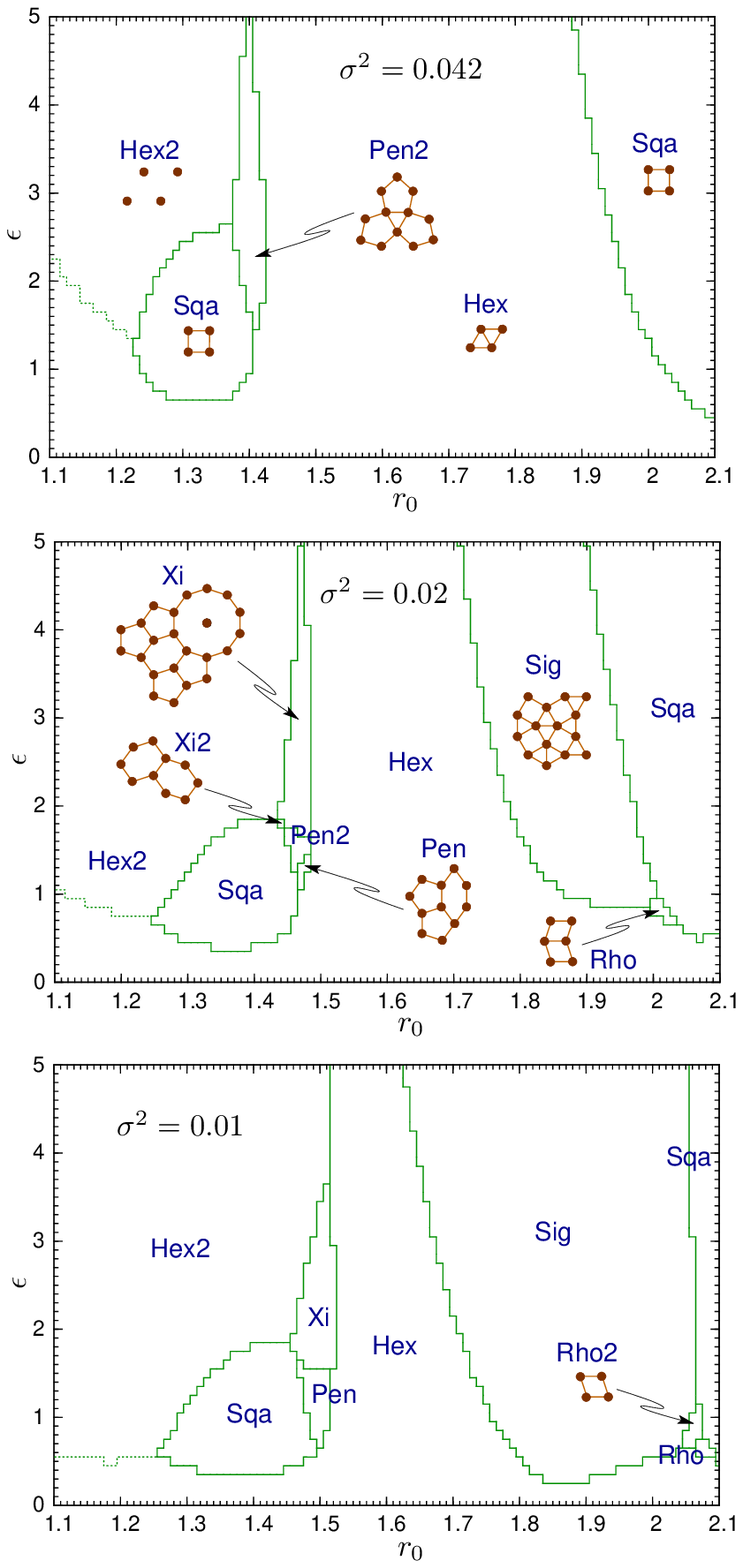}
\caption{Phase diagram of the two-dimensional LJG system for three values of
  $\sigma^2$. For each crystal the decoration of a unit cell is shown.}
\label{fig:2}
\end{figure}

Next, we calculate the $T=0$, $p=0$ phase diagram in the
$r_0$-$\epsilon$-$\sigma^2$ parameter space via annealing simulations and
structural relaxation. For details, we refer to \cite{Engel2007}, where the
same procedure has been applied. The result is depicted in Fig.~\ref{fig:2}. A
wide second minimum ($\sigma^2=0.042$) stabilizes mostly simple phases. There
are two hexagonal lattices (Hex, Hex2), which are connected continuously along
small $\epsilon$-values. A rapid increase of the lattice constant is observed
at the dashed line. Additionally, the square lattice (Sqa) and a phase built
from deformed pentagons and triangles (Pen2) are found.

Further structures appear for $\sigma^2=0.02$: a complicated phase with
decagons (Xi), a related one with flattened honeycombs (Xi2), a phase with
pentagons and hexagons (Pen), the two-dimensional sigma phase (Sig), and
alternating rhombs (Rho). In the case of $\sigma^2=0.01$, parallel rhombs
(Rho2) are stable. We remark that the phase diagram differs slightly from the
one published in \cite{Engel2007} due to a higher precision of the relaxation:
Sig2 and Sig3 are now unstable, whereas Xi2 and Pen2 appear as new phases. In
total, nine ground states have been discovered, among them five complex
crystals (Tab.~\ref{tab:1}).

\begin{table}
  \centering
  \begin{tabular}{ccclccc}\hline\hline
    Phase & Dim & Density & Lattice constants & Particles/u.c. & Symmetry &
    References\\\hline
    Xi     & 2 & 0.76 & $a=4.24$, $\alpha=72^{\circ}$ &13 &cmm &
    \cite{Engel2007}\\
    Xi2    & 2 & 0.80 & $a=1.62$, $b=3.08$ & 4 & pgg  & \cite{Engel2007}\\
    Sig    & 2 & 1.07 & $a=2.73$           & 8 & p4g  & \cite{Quandt1999,
      Chen2005, Hayashida2007}\\ 
    Pen    & 2 & 0.90 & $a=2.62$, $b=4.25$ &10 & cmm  & \cite{Skibinsky1999,
      Jagla1998}\\
    Pen2   & 2 & 0.99 & $a=2.41$           & 5 & p31m & \cite{Engel2007}
    \\\hline
    sig    & 3 & 1.01 & $a=3.86$, $b=2.00$ &30 & P4$_2$/mnm &
    \cite{Donohue1982}  \\\hline\hline
  \end{tabular}
  \caption{The complex crystals observed in the LJG system. The
    nearest-neighbor distance is equal 1.}
  \label{tab:1}
\end{table}

The sequence of phases in Fig.~\ref{fig:2} follows the local order analysis in
Fig.~\ref{fig:1}. Hence, the phase diagram can be understood as the result of
stabilizing regular polygons. We assume that the behavior is universal in the
sense that the phase diagram looks similar for most two-scale interaction
potentials.

\begin{figure}
\centering \includegraphics[width=0.7\textwidth]{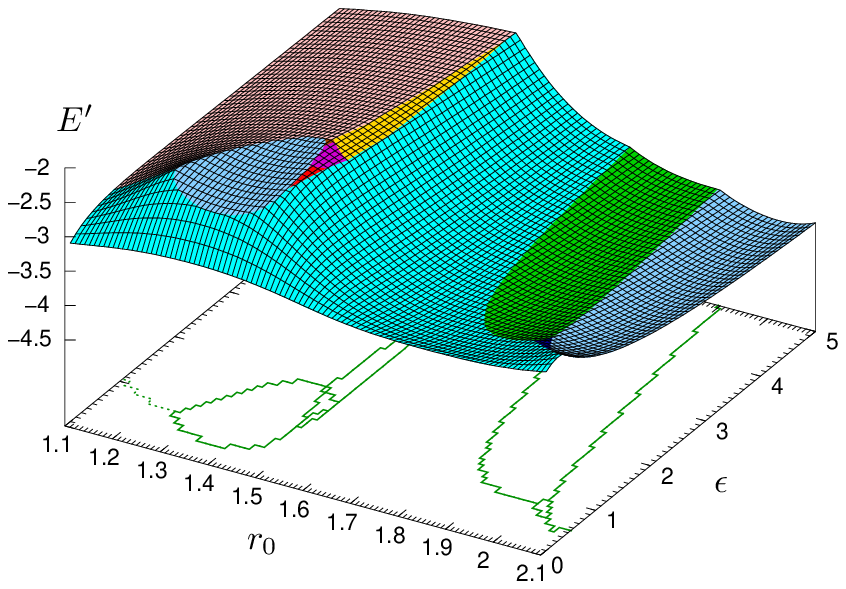}
\caption{Rescaled ground state energy $E'$ for $\sigma^2=0.02$. Maximum and
  minimum are located in the stability regions of Sqa.}
\label{fig:3}
\end{figure}

It is interesting to study the ground state energy per particle, $E$, as a
function of the potential parameters. In Fig.~\ref{fig:3} the rescaled energy
$E'(r_0,\epsilon)=E/(\epsilon+1)$ is plotted. The rescaling enforces
convergence in the limit $\epsilon\rightarrow\infty$. For a given phase the
energy has a minimum near $r_0$ defined by (\ref{eq:2}) and
(\ref{eq:3}). Sometimes the phase is not stable at its minimum energy. As seen
in Fig.~\ref{fig:3}, the stability regions of Pen/Pen2 and Xi/Xi2 are shifted
to lower $r_0$ values due to competition with Hex.

An important aspect of the two-dimensional LJG system is the fact, that all
phases can actually be grown as single crystals in simulations, if the cooling
time is slow enough.  When heating the system up, the complex crystal often
reversibly transform into quasicrystals. This is the case for Xi
\cite{Engel2007}, but also for Pen/Pen2 and Sig \cite{Engel2008b}, and means
that these phases are not accessible from the melt, but have to form via a
solid-solid transformation from the quasicrystal \cite{Engel2008}.

\section{The sigma phase and a glass in three dimensions}

In this section, first results for the three-dimensional LJG system are
presented. Its full $T=0$, $p=0$ phase diagram is not known. Nevertheless as
we show now, the LJG potential allows to grow at least one complex crystal in
simulations. We ran 11 simulations with $\epsilon=1.8$, $\sigma^2=0.02$, and
$r_0\in[1.0,2.0]$, $\Delta r_0=0.1$. Standard molecular dynamics (MD) with a
Nos\'e-Hoover thermo-/barostat and periodic boundary conditions were
applied. The cubic simulation box contained 2744 particles with an initially
amorphous configuration. We searched for temperatures close to, but below the
melting point for rapid crystallization.

We were successful to achieve crystallizations for all $r_0$ values except
1.4. After a few $10^5$ MD steps, six different crystals have been grown: fcc
(cF4, $r_0=1.0$), hcp (hP2, 1.6), bcc (cI2, 1.1-1.2, 1.7-1.9), shl (hP1, 1.3),
pzt (tP5, 2.0), sig (tP30, 1.5). Here, shl is the simple hexagonal lattice
(prototype: $\gamma$-HgSn$_{6-10}$, disordered), pzt the tetragonal perovskite
structure (PbZr$_\text{x}$Ti$_{1-\text{x}}$O$_3$), and sig the sigma-phase
($\beta$-uranium).

It was observed that the particles cannot relax effectively with periodic
boundaries, since point defects and dislocations do not heal out. Therefore,
the simulations were repeated with open boundaries, i.e.\ the particles form a
solid sphere floating in vacuum. The structures are the same, but this time
single crystals were obtained. The possibility to simulate at zero pressure is
an advantage of the LJG potential. For the Dzugutov potential, external
pressure usually has to be applied, because of the repulsive bump in the
interaction.

For $r_0=1.5$, the sigma phase was observed. To study its formation in detail,
we simulated a large system of $15\,625$ particles at $T=1.7$ (as usual
$k_B=1$), which is about $90\%$ of the melting temperature. After less than
$10^5$ MD steps, local crystallization started and a polycrystal with
dodecagonal quasicrystallites appeared. Ordering proceeded in two steps: first
to a bicrystal at $10^6$ steps and then to a single crystal at $2\cdot 10^6$
steps. At the same time, rearrangements within each crystallite transformed
the quasicrystal state into the sigma phase. After $3\cdot 10^6$ steps the
system was relaxed to $T=0$.

\begin{figure}
\centering \includegraphics[width=\textwidth]{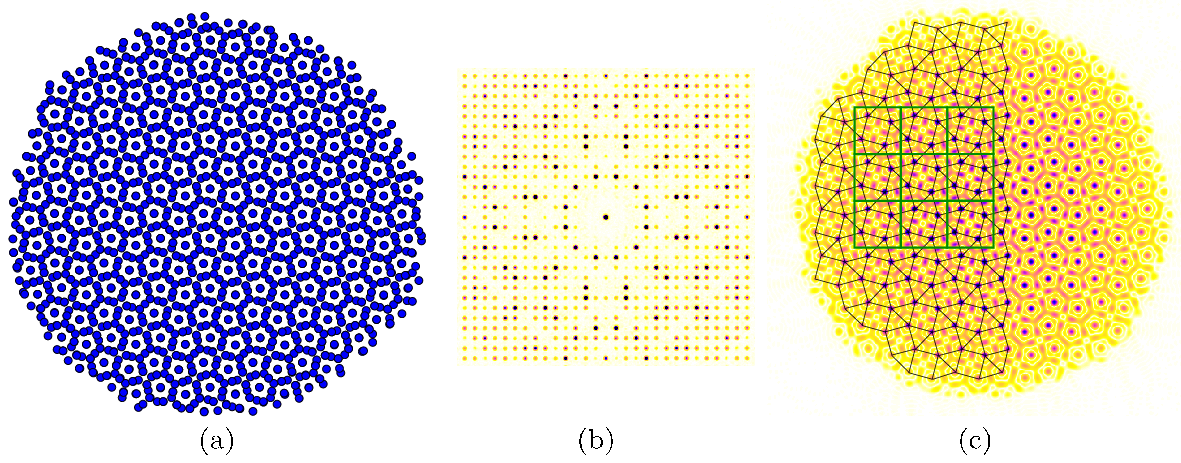}
\caption{Single crystal of the sigma phase grown with molecular dynamics using
  $15\,625$ particles. (a)~Projection along the four-fold symmetry
  axis. (b)~Diffraction image. (c)~Fourier-filtered structure image with
  tiling superimposed.}
\label{fig:4}
\end{figure}

Fig.~\ref{fig:4}(a) shows the configuration of the particles projected along
the four-fold axis. The width of the Bragg peaks in the diffraction image
(Fig.~\ref{fig:4}(b)) is determined by the system size. By selecting only the
twelve inner strong reflections for an inverse Fourier transform, we obtain a
Fourier filtered structure image (Fig.~\ref{fig:4}(c)). It allows to easily
detect the underlying tiling. Cylindrical columns with pentagonal shape
forming squares and triangles in a ratio 1:2 are characteristic for the sigma
phase.

We note that the sigma phase is unstable at $T=0$, because fcc has a lower
energy than sig. To test the stability at $T=1.7$, we initiated a simulation
of a fcc crystal in contact with a sigma crystal. Quickly, the whole system
transformed into a single fcc phase. Hence, the sigma phase is also not
stabilized entropically. The reason for its appearance in our simulations is
the huge nucleation radius of fcc. The local order of sig is much closer to
the melt. Further details will be reported elsewhere.

In the case $r_0=1.4$, all attempts to crystallize the system failed. One
reason is the comparably low melting temperature: the ground state fcc melts
around $T=0.9$. In various MD runs over several $10^6$ steps in the range
$0.5\leq T\leq 0.9$ no nucleation was observed. We conclude that the choice
$r_0=1.4$ is interesting for studying a monatomic glass. Our results indicate
that this LJG glass is more resistant against crystallization than the
Dzugutov glass \cite{Dzugutov1992}, which forms the dodecagonal quasicrystal
rather quickly \cite{Dzugutov1993, Roth2000, Keys2007}.

\section{Discussion}

How does the LJG system compare to experiments? In metals, multi-body terms
can only be neglected in a first approximation. Effective potentials often
have Friedel oscillations, which are mimicked by double-wells. It should be
kept in mind that fixed pair interactions between all atoms might be not
applicable in complex phases, since the atoms are found in different local
environments. Furthermore, the interaction is expected to change during
crystallization.

Isotropic pair potentials are more applicable to macromolecular self-assembly,
because the molecules as a whole interact almost classically. It has been
suggested \cite{Lifshitz2007} that their complex arrangements originates from
the competition of two length scales, which appear due to soft repulsion and
strong interpenetration. The micelles forming the dodecagonal quasicrystal in
\cite{Zeng2004} have two natural length scales: the inner one corresponds to
the backbone of the dendrons and the outer one to the end of the tethered
chains. Our simulations suggest that the ideal ratio of the scales is close to
2:3.

The cylindrical phases (T-shaped molecules and ABC-star polymers) can be
understood as two-dimensional tilings with an effective interaction within the
plane. Although not isotropic anymore, the particles still have two length
scales. The tilings observed so far \cite{Matsushita2007, Tschierske2007}
consist of hexagons, squares, and triangles only. If it will be possible to
stabilize pentagons or decagons, then the phases Pen/Pen2, Xi/Xi2 might
appear.

We finish with a challenge: Can monodisperse icosahedral quasicrystals be
grown in simulation or experiment? As of today, none have been found.

\section{Acknowledgment}

This work was funded by the Deutsche Forschungsgemeinschaft (Tr 154/24-1) and
the Japan Society for the Promotion of Sciences.

\end{document}